\documentclass[runningheads,a4paper]{llncs}
\usepackage{xcolor}
\usepackage{graphicx}
\usepackage{epstopdf,subfig,epsfig,float}

\usepackage{amssymb}
\usepackage{chngpage} 
\usepackage{multirow} 
\usepackage{enumitem} 

\usepackage{url}

\let\oldbibliography\thebibliography
\renewcommand{\thebibliography}[1]{\oldbibliography{#1}
\setlength{\itemsep}{0pt}} 
\urldef{\mailsa}\path|xiangju.qin@ucdconnect.ie, {derek.greene,padraig.cunningham}@ucd.ie|

\begin{document}

\mainmatter  

\title{A Latent Space Analysis of Editor Lifecycles in Wikipedia}

\titlerunning{Latent Space Analysis of Editor Lifecycles}

%
%
\author{Xiangju Qin%
\and Derek Greene
\and P\'{a}draig Cunningham}
\authorrunning{X. Qin, D. Greene, P. Cunningham}

\institute{School of Computer Science \& Informatics, University College Dublin\\
\mailsa\\
}

%
%

\toctitle{Lecture Notes in Computer Science}
\tocauthor{Authors' Instructions}
\maketitle

\begin{abstract}

Collaborations such as Wikipedia are a key part of the value of the modern Internet. At the same time there is concern that these collaborations are threatened by high levels of member turnover. In this paper we borrow ideas from topic analysis to editor activity on Wikipedia over time into a latent space that offers an insight into the evolving patterns of editor behavior. This latent space representation reveals a number of different categories of editor (e.g. content experts, social networkers) and we show that it does provide a signal that predicts an editor's departure from the community. We also show that long term editors gradually diversify their participation by shifting edit preference from one or two namespaces to multiple namespaces and experience relatively soft evolution in their editor profiles, while short term editors generally distribute their contribution randomly among the namespaces and experience considerably fluctuated evolution in their editor profiles.

\end{abstract}

\vspace{-0.15in}
\section{Introduction}
\vspace{-0.1in}
With the popularity of Web 2.0 techniques, recent years have witnessed an increasing population of online peer production communities which rely on contributions from volunteers to build software and knowledge artifacts, such as Wikipedia, OpenStreetMap and StackOverflow. The growing popularity and importance of these communities requires a better understanding and characterization of user behavior so that the communities can be better managed, new services delivered, challenges and opportunities detected. For instance, by understanding the general lifecycles that users go through and the key features that distinguish different user groups and different life stages, we can: (i) predict whether a user is likely to abandon the community; (ii) develop intelligent task routing software to recommend tasks to users within the same life-stage. Moreover, the contribution and social interaction behavior of contributors plays an essential role in shaping the health and sustainability of online platforms.

Recent studies have approached the issue of modeling user lifecycles (also termed as user profiles or user roles) in online communities from different perspectives. Such studies have so far focused on separate or a combination of user properties, such as information exchange behavior in discussion forums \cite{Chan2010}, social and/or lexical dynamics in online platforms \cite{Danescu-Niculescu-Mizil2013,Rowe2013}, and diversity of contribution behavior in Q\&A sites \cite{Furtado2013}. These studies generally employed either principle component analysis and clustering analysis to identify user profiles \cite{Chan2010,Furtado2013} or entropy measure to track social and/or linguistic changes throughout user lifecycles \cite{Danescu-Niculescu-Mizil2013,Rowe2013}. While previous studies provide insights into community composition, user profiles and their dynamics, they have limitations either in their definition of lifecycle periods (e.g., dividing each user's lifetime using a fixed time-slicing approach \cite{Danescu-Niculescu-Mizil2013} or a fixed activity-slicing approach \cite{Rowe2013}) or in the expressiveness of user lifecycles in terms of the evolution of expertise and user activity for users and the communities over time. Specifically, they fail to capture a mixture of user interests over time.

On the other hand, in recent years, there has been significant advances in topic models which develop automatic text analysis models for discovering latent structures from time-varying document collections. In this paper we present a latent space analysis of user lifecycles in online communities specifically Wikipedia. Our contributions are as follows:
\vspace{-.6em}
\begin{itemize}
	\item We model the lifecycles of users based on their activity over time using topic modeling, thus complementing recent work (e.g. \cite{Danescu-Niculescu-Mizil2013,Furtado2013,Rowe2013}).
	\item This latent space analysis reveals a number of different categories of editor (e.g. content experts, social networkers) and offers an insight into the evolving patterns of editor behavior. 
	\item We find that long term and short term users have very different profiles as modeled by their activity in this latent representation.
	\item We show that the patterns of change in user activity can be used to make predictions about the user's membership in the community.
\end{itemize}

The rest of this paper is organized as follows. The next section provides a brief review of related work. In Section 3, we provide an overview of dynamic topic models and explain data collection. Next, we present results about latent space analysis of editor lifecycles in Wikipedia, followed by results about churn prediction and conclusions.

\vspace{-0.1in}
\section{Related Work}
\vspace{-0.1in}
Over the last decade, the study of investigating and modeling the changes in user behavior in online communities has gripped researchers. Such studies have been conducted in varied contexts, including discussion forums \cite{Chan2010}, Wikipedia \cite{Panciera2009,Welser2011}, beer rating sites \cite{Danescu-Niculescu-Mizil2013}, Q\&A sites \cite{Rowe2013,Furtado2013} and other wikis. Chan {\it et al.} \cite{Chan2010} presented an automated forum profiling technique to capture and analyze user interaction behavior in discussion forums, and found that forums are generally composed of eight behavior types such as popular initiators and supporters. Welser {\it et al.} \cite{Welser2011} examined the edit histories and egocentric network visualizations of editors in Wikipedia and identified four key social roles: substantive experts, technical editors, vandal fighters, and social networkers. Panciera {\it et al.} \cite{Panciera2009} studied the contribution behaviors of long-term editors and newcomers in Wikipedia and their changes over time, and found significant difference between the two groups: long-term editors start intensely, tail off a little, then maintain a relatively high level of activity over the course of their career; new users follow the same trend of the evolution but do much less work than long-term editors throughout their lifespans. The studies mentioned provide insights about contributor behavior at a macro level, but are limited in capturing the change of behavior at a user level.

Danescu-Niculescu-Mizil {\it et al.} \cite{Danescu-Niculescu-Mizil2013} examined the linguistic changes of online users in two beer-rating communities by modeling their term usages, and found that users begin with an innovative learning phase by adopting their language to the community but then transit into a conservative phase in which they stop changing the language. Rowe \cite{Rowe2013} modeled how the social dynamics and lexical dynamics of users changed over time in online platforms relative to their past behavior and the community-level behavior, mined the lifecycle trajectories of users and then used these trajectories for churn prediction. Based on the diversity, motivation and expertise of contributor behaviors in five Q\&A sites, Furtado {\it et al.} \cite{Furtado2013} examined and characterized contributor profiles using hierarchical clustering algorithm and K-means algorithm, and found that the five sites have very similar distributions of contributor profiles. They further identified common profile transitions by a longitudinal study of contributor profiles in one site and found that although users change profiles with some frequency, the site composition is mostly stable over time. The aforementioned works provide useful insights into community composition, user profiles and their dynamics, but they are limited either in their definition of lifecycle periods (e.g., dividing each user's lifetime using a fixed time-slicing approach \cite{Danescu-Niculescu-Mizil2013} or a fixed activity-slicing approach \cite{Rowe2013}) or in the expressiveness of user profiles in terms of the evolution of expertise and user activity for users and the communities over time \cite{Furtado2013}. Specifically, they fail to capture a mixture of user interests over time.

In recent years, there has been an increasing interest in developing automatic text analysis models for discovering latent structures from time-varying document collections. Blei and Lafferty \cite{Blei2006} presented a dynamic topic model (DTM) which utilizes state space models to link the word distribution and popularity of topic over time. Wang and McCallum \cite{Wang2006} proposed the topics over time model which employs a beta distribution to capture the evolution of topic popularity over timestamps. Ahmed and Xing \cite{Ahmed2010} introduced the infinite dynamics topic models (iDTM) which can adapt the number of topics, the word distributions of topics, and the topics' popularity over time. Based on topic models, Ahmed and Xing (2011) proposed a time-varying user model (TVUM) which models the evolution of topical interests of a user while allowing for user-specific topical interests and global topics to evolve over time. Topic modeling plays a significant role in improving the ways users search, discover and organize web content by automatically discovering latent semantic themes from a large and otherwise unstructured collection of documents. Moreover, topic modeling algorithms can be adapted to many types of data, such as image datasets, genetic data and history user activity in computational advertising. However, to our knowledge, there exists no attempt to understand how users develop throughout their lifecycles in online communities from the perspective of topic modeling.

This paper complements the previous works by characterizing the evolution of user activity in online communities using dynamic topic modeling, examines the patterns of change in users as modeled by their activity in the latent representation and demonstrates the utility of such patterns in predicting churners.

\vspace{-0.2in}
\section{Model Editor Lifecycles}
In this section, we provide a brief overview about dynamic topic models that we will use to model the lifecycle of Wikipedia users and introduce how we collect the data for evaluation. 
\vspace{-0.1in}
\subsection{Dynamic Topic Model}
The primary goal of this study is to apply topic models on the evolving user activity collections in order to identify the common work archetypes\footnote{Common work archetypes refer to the types of contribution that users make in online platforms, e.g., answering questions in Q\&A sites, editing main pages in Wikipedia.} and to track the evolution of common work archetypes and user lifecycles in online communities. For this purpose, we employ an LDA based dynamic topic model proposed by Blei and Lafferty \cite{Blei2006}, in which the word distribution and popularity of topics are linked across time slices using state space models. First, we review the generative process of the LDA model \cite{Blei2003}, in which each document is represented as a random mixture of latent topics and each topic is characterized by a multinomial distribution over words, denoted by Multi($\beta$). The process to generate a document $d$ in LDA proceeds as follows:
\vspace{-0 mm}
\begin{enumerate}
	\item Draw topic proportions $\theta_{d}$ from a Dirichlet prior: $\theta_{d}|\alpha\sim Dir(\alpha)$.
	\item For each word
	\begin{enumerate}
		\item Draw a topic assignment from $\theta_{d}$: $z_{di}|\theta_{d}\sim Mult(\theta_{d})$.
		\item Draw a word $w_{di}$: $w_{di}|z_{di},\beta\sim Mult(\beta_{z_{di}})$. 
	\end{enumerate}	 
\end{enumerate}
\vspace{-0 mm}
Where $\alpha$ is a vector with components $\alpha_{i}>0$; $\theta_{d}$ represents a topic-mixing vector for document $d$ that samples from a Dirichlet prior (i.e. Dir($\alpha$)), each component (i.e. $z_{di}$) of $\theta_{d}$ defines how likely topic $i$ will appear in $d$; $\beta_{z_{di}}$ represents a topic-specific word distribution for topic $z_{di}$.

LDA is not applicable to sequential models for time-varying document collections for its inherent features and defects: (1) Dirichlet distributions are used to model uncertainty about the distributions over words and (2) the document-specific topic proportions $\theta$ are drawn from a Dirichlet distribution. To remedy the first defect, Blei and Lafferty \cite{Blei2006} chained the multinomial distribution of each topic $\beta_{t,k}$ in a state space model that evolves with Gaussian distributions, denoted as follows:
\begin{eqnarray}
\label{formula:Gaussian_dist_topic}
   \beta_{t,k}|\beta_{t-1,k}\sim N(\beta_{t-1,k},\sigma^{2}I)
\end{eqnarray}
To amend the second defect, the same authors employed a logistic normal with mean $\alpha$ to capture uncertainty over proportions and used the following dynamic model to chain the sequential structure between models over time slices \cite{Blei2006}:
\begin{eqnarray}
\label{formula:draw_doc_topic_dist}
   \alpha_{t}|\alpha_{t-1}\sim N(\alpha_{t-1},\sigma^{2}I)
\end{eqnarray}

More details on the generative process of a dynamic topic model for a sequential corpus can be found in \cite{Blei2006}. Note that documents generated using the DTM will have a mixture of topics. In Wikipedia, this indicates that a user has diverse edit interests and edits multiple namespaces in a specific time period.

\vspace{-0.1in}
\subsection{Data Collection}
\vspace{-0.05in}
In Wikipedia, the pages are subdivided into `namespaces'\footnote{http://en.wikipedia.org/wiki/Wikipedia:Namespace} which represent general categories of pages based on their function. For instance, the article (or main) namespace is the most common namespace and is used to organize encyclopedia articles. Users can make edits to any namespace based on their edit preference. The amount of edits across all the namespaces can be considered as work archetypes. A namespace can be considered as a `term' in the vector space for document collections, the number of edits to that namespace is analogous to word frequency. A user's edit activity across different namespaces in a time period can be regarded as a `document'. One motivation of this study is to identify and characterize the patterns of change in user edit activity over time in Wikipedia. For this purpose, we parsed the August 2013 dump of English Wikipedia\footnote{http://dumps.wikimedia.org/enwiki/20130805/}, collected the edit activity of all registered users, then aggregated the edit activity of each user on a quarterly basis  (measured in 3-month period). In this way, we obtained a time-varying dataset consisting of the quarterly editing activity of all users from the inception of Wikipedia till August 5th, 2013. The statistics about the complete dataset are as follows: 51 quarters; 28 namespaces (or features); 5,749,590 unique registered users; 10,336,278 quarterly edit observations for all users\footnote{The first quarter for each user is the same, i.e., the one where Wikipedia was founded. For a specific quarter, a user had observation only if that user had edit activity.}. An example of our dataset is as follows:
\vspace{-8 mm}
\begin{table}[!htb]
\caption{A simple example of the dataset}
\label{}
\renewcommand{\arraystretch}{1.2}
{\fontsize{8.0pt}{8.0pt}\selectfont
\begin{tabular}{l|l|l|l|l|l|l|l|l}
\hline
Uname & quarter & articles & article talk & wikipedia & wikipedia talk & cat temp port pages & user & user talk \\ \hline
User A & \multicolumn{1}{c|}{1} & \multicolumn{1}{c|}{650} & \multicolumn{1}{c|}{233} & \multicolumn{1}{c|}{2} & \multicolumn{1}{c|}{299} & \multicolumn{1}{c|}{33} & \multicolumn{1}{c|}{0} & \multicolumn{1}{c}{81} \\ \hline
\end{tabular}
}
\end{table}
\vspace{-5 mm}

\begin{figure}[!htb]
{\fontsize{10.0pt}{10.0pt}\selectfont
\begin{center}
  \subfloat{\includegraphics[width=1.0\textwidth]{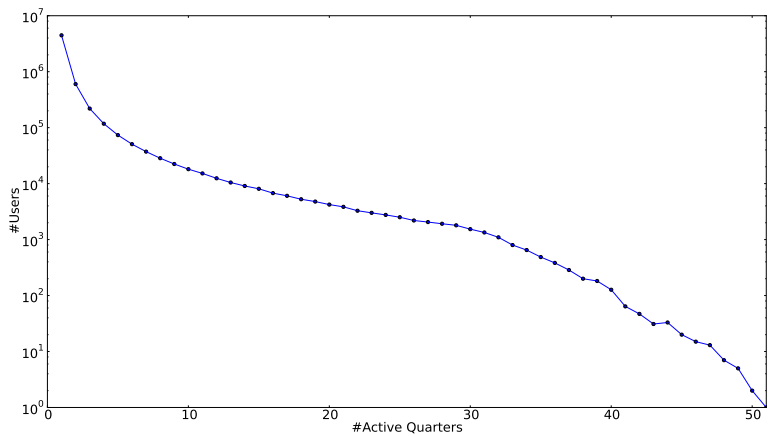}}
\vspace{-.8em}
\caption{Statistics about the lifespans of all registered users in Wikipedia. X-axis represents the number of active quarters, y-axis represents the number of users who stayed active for a specific number of quarters, and is ploted in log10 scale.}
\vspace{-1.2em}
\label{fig:Registered_users_active_stats}
\end{center}
}
\end{figure}

Figure \ref{fig:Registered_users_active_stats} plots the statistics about the number of active quarters for all registered users. We observe that an overwhelming number of users (i.e. 4,468,352 out of  5,749,590) stayed active for only one quarter; the figures for 2, 3 and 4 quarters are 600,417, 219,455 and 117,698, respectively. One obvious trend is that the number of users who stayed active for longer time periods is becoming smaller and smaller, indicating that Wikipedia experiences high levels of member withdrawal. To avoid a bias towards behaviors most dominate in communities with larger user bases, following Furtado {\it et al.} \cite{Furtado2013}, we randomly selected 20\% of users who stayed active for only one quarter, included these users and those who were active for at least two quarters into our dataset, generating a time-varying dataset with 2,162,978 unique users and 6,661,973 observations.
\vspace{-5 mm}

\vspace{-0.1in}
\section{Editor Lifecycles as Released by Shift in Participation}
In this section, we present the analysis of editor lifecycles from two perspectives: (1) community-level evolution; and (2) user-level evolution for group of editors. From the analysis, we identify some basic features that are useful for predicting how long will a user stay active in an online community (Section \ref{sec:Churn_prediction}).

\subsection{Community level change in lifecycle}
\begin{figure}[!htb]
\baselineskip=12pt
{\fontsize{10.0pt}{10.0pt}\selectfont
\begin{center}
  \subfloat[User role ``All-round contributors"]{\includegraphics[width=.85\textwidth]{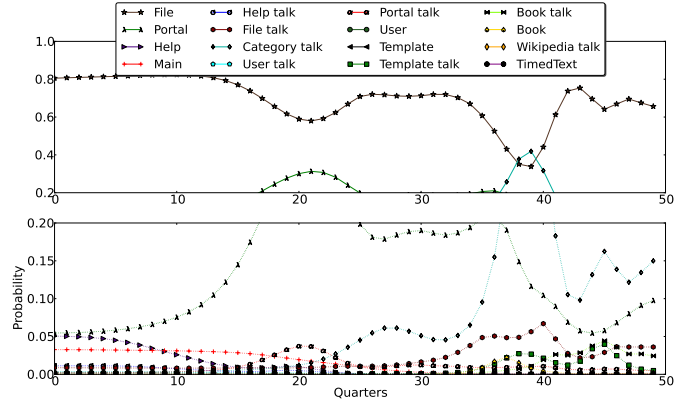}}\quad
  \subfloat[User role ``Technical experts"]{\includegraphics[width=.85\textwidth]{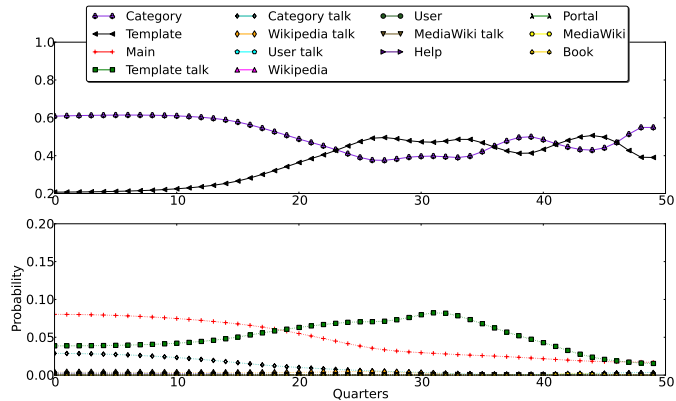}}\\
\vspace{-.8em}  
\caption{Example of the evolution of the common user roles. X-axis corresponds to quarters, y-axis indicates the probability of a namespace appearing in a user role profile. Each curve represents the evolution of the probability of a namespace appearing in the profile. Note that the scales are different in two parts of y-axis.}
\vspace{-1.2em}
\label{fig:user_role_evolution}
\end{center}
}
\end{figure}

Online communities experience dynamic evolution in terms of a constantly changing user base and the addition of new functionalities to maintain vitality. For instance, online platforms like Wikipedia generally experience high levels of member withdrawal, with 60\% of registered users staying only a day. Wikipedia introduced social networking elements to MediaWiki in order to attract and retain user participation\footnote{http://strategy.wikimedia.org/wiki/Attracting\_and\_retaining\_participants}, which is believed to increase user edit activity. As such, to understand what shape these communities, it is essential to take into account both dimensions of evolution. The dynamic topic models can accommodate both aspects of evolution, which provides valuable insights about the development of community and its users across time. To analyze the evolution of time-varying user edit activity, we ran the DTM software\footnote{Available at: \url{http://www.cs.princeton.edu/~blei/topicmodeling.html}} released by Blei and Lafferty \cite{Blei2006} with default hyperparameters and different numbers of topics $k\in$ [5,7,8,9,10,15]\footnote{We analyzed the runs with different $k$, found that the run with 7 topics provides more interpretable and expressiveness results in terms of interpretation and overlapping between different topics. The results are reported in Table \ref{Tab:User_Role_Summary} and Figure \ref{fig:user_role_evolution}.}.

\vspace{-5 mm}
\begin{table}[!htbp]
\renewcommand{\arraystretch}{1.2}
\vspace{-1.2em}
\caption{Summary of common user roles. Dominant features are in bold font.}
\vspace{-.5em}
\label{Tab:User_Role_Summary}
{\fontsize{8.0pt}{8.0pt}\selectfont
\begin{center}
\begin{tabular}{c|l|c}
\hline
Id & Name & Edit Namespaces (Sequence indicates the importance) \\ \hline
1& Social Networkers & \emph{\textbf{user talk}, main, Wikipedia talk} \\ \hline
\multirow{2}{*}{2}& \multicolumn{1}{l|}{Content Proposal;} & \multirow{2}{*}{\emph{\textbf{article talk}, \textbf{main}, user talk, Wikipedia talk}} \\
& conversial coordinator &  \\ \hline

3& Content Experts & \emph{\textbf{main}} \\ \hline
4& All-round contributors & \emph{\textbf{file}, \textbf{portal}, \textbf{category talk}, main, portal talk, book talk, template talk} \\ \hline
5& Admin related roles I & \emph{\textbf{user}, main, article talk, user talk} \\ \hline
6& Technical experts & \emph{\textbf{category}, \textbf{template}, main, template talk, category talk} \\ \hline
7& Admin related roles II & \emph{\textbf{Wikipedia}, \textbf{Wikipedia talk}, main, user talk, article talk, user} \\ \hline
\end{tabular}
\end{center}
}
\end{table}
\vspace{-5 mm}

\textbf{Summary of user roles.} Table \ref{Tab:User_Role_Summary} presents a summary of the common user roles identified by DTM \cite{Blei2006}. It is obvious from Table \ref{Tab:User_Role_Summary} that, each user role is defined by different combinations of namespaces. For instance, user role ``Social Networkers" is dominated by \emph{user talk} namespace, then \emph{main} namespace and \emph{Wikipedia talk} namespace, indicating that users who are assigned to this group spend most of their time in Wikipedia interacting with other users; user role ``All-round contributors" is dominated by \emph{file, portal, category talk, main} and other namespaces, as suggested by the large amount of namespaces, this group of users contribute to a diversity of namespaces and hence the name for this user role; user role ``Admin related roles I" and ``Admin related roles II" mainly relate to maintenance, management and organization aspects of Wikipedia. Note that different user roles correspond to different common work archetypes.

\textbf{Community-level change in user roles.} As users stay long enough in Wikipedia, they tend to shift their participation by changing their edits to other namespaces. Moreover, the addition of new functionalities also encourages shift in user participation. Figure \ref{fig:user_role_evolution} visualizes the evolution of two user roles at an aggregate level, from which we make several observations. First, in Figure \ref{fig:user_role_evolution}(a), the probabilities fluctuate over time, with namespaces emerged in some timestamp. For example, \emph{category talk} namespace gradually emerges as a feature from the 21st quarter, reaches its peak at the 38th quarter, followed by a decrease afterwards. The emergence of namespaces may well be corresponding to new elements being added to Wikipedia, which promotes shift in user participation. Second, by contrast, in Figure \ref{fig:user_role_evolution}(b), the probabilities experience a more soft evolution, with \emph{category} and \emph{template} namespaces dominated the user role profile. Last, corresponding to the emergence of namespace, some namespaces appear as features in profiles with low probabilities in the early lifespan, but disappear from the profiles in the late lifespan. For instance, in Figure \ref{fig:user_role_evolution}(b), \emph{category talk} namespace appears as a feature in the first 20 quarters with a probability of 0.03, but disappears from this quarter onwards.

\vspace{-5 mm}

The other user roles evolve in similar ways as those in Figure \ref{fig:user_role_evolution}, but are omitted due to space limitation. The evolution of user roles captures the overall changes in editor profiles due to the addition of new functionalities and shifts in user participation over time, but it provides little insight about what is the general trajectory of a user's participation as the user transitions from being a newcomer to being an established member of the community. We explore this question in the next section. 

\vspace{-0.15in}
\subsection{User Lifecycle}
\vspace{-0.08in}
We now move on to examining how group of users evolve throughout their lifecycle periods. Understanding how individual users develop over time in online communities makes it possible to develop techniques for important tasks such as churn prediction, as we will show through experiments in Section \ref{sec:Churn_prediction}.

\textbf{Clustering of Wikipedia editor profiles.} Different from previous studies which analyze user lifecycle on a medium size dataset (with about 33,000 users \cite{Danescu-Niculescu-Mizil2013,Furtado2013,Rowe2013}), our analysis is based on a large dataset with approximately half a million users\footnote{For the clustering and churn analysis, we only considered editors who were active for at least 4 quarters, resulting in 455233 editors. The data for the 51st quarter (01/07--30/09, 2013) was incomplete and excluded from the analysis.}. Different users are more likely to follow a slightly or totally different trajectories in their lifecycle. As a result, it is very unlikely for us to employ one single model to perfectly fit the development of lifecycle for all users, as Rowe \cite{Rowe2013} chose linear regression model to characterize the development of user properties. On the other hand, Non-negative Matrix Factorization (NMF) clustering is an efficient clustering approach for large scale dataset. In this analysis, we employ the NMF approach to cluster Wikipedia editor profiles, and identify the general trajectories for group of editors.

To perform NMF clustering on the dataset, we restructured the editor profiles as follows: for each editor, we combined all her quarterly profile assignments obtained by DTM into one record, with each dimension representing the probability that the editor of a specific quarter was assigned to a user role. If an editor had no edit activity in a quarter and thus had no profile assignment, we used missing data (i.e. 0 values) to represent the corresponding dimensions. 

We clustered the Wikipedia editor profile data as follows. Firstly, we constructed a  $455233 \times 350$ profile-feature matrix, where each row corresponds to a different editor profile. Rows were subsequently L2 normalized to ensure that each profile vector had unit length. To cluster this matrix efficiently, we use the fast alternating least squares variant of NMF introduced by Lin \cite{lin07gradient}. To produce deterministic results and avoid a poor local minimum, we used the Non-negative Double Singular Value Decomposition (NNDSVD) strategy \cite{bout08headstart} to choose initial factors for NMF. We ran this process for different numbers of clusters $K \in [4,10]$, and inspected both the resulting coefficient matrix (\emph{i.e.}\,the profile clusters) and basis vector matrix (\emph{i.e.}\,the feature clusters) for each value of $k$. Finally, to produce disjoint clusters, where each profile is assigned to a single cluster, we discretized the coefficient matrices. Table \ref{Tab:stats_for_NMF_clustering} presents a summary of NMF clustering on the editor profile data\footnote{We applied NMF clustering on the dataset with $K \in [4,10]$, and found that the clustering with $K$=10 provides a more reasonable grouping of editors.}.

\begin{table}[!htb]
\renewcommand{\arraystretch}{1.2}
\vspace{-1.2em}
\caption{Summary for NMF clustering of editor profiles (N=455233)}
\label{Tab:stats_for_NMF_clustering}
{\fontsize{8.0pt}{8.0pt}\selectfont
\begin{center}
\begin{tabular}{c|c|c|c|c|c|c|c|c|c}
\hline
\multirow{2}{*}{cluster\_id} & \multirow{2}{*}{\#editors} & \multirow{2}{*}{Fraction} & \multicolumn{ 4}{c|}{Active quarter statistics} & \multicolumn{ 2}{c|}{Dominant features} & \multirow{2}{*}{\#admins} \\ \cline{ 4- 9}
 &  &  & \multicolumn{1}{l|}{Min} & \multicolumn{1}{l|}{Max} & \multicolumn{1}{l|}{Median} & \multicolumn{1}{l|}{Avg} & Quarters & \multicolumn{1}{l|}{User roles} & \\ \hline
1 & 9673 & 0.021 & 7 & 20 & 9 & 8.633 & 1--8 & 3, 2, 5 & 12 \\ \hline
2 & 34348 & 0.075 & 13 & 42 & 19 & 19.279 & 8--17 & 3 & 317 \\ \hline
3 & 15812 & 0.035 & 20 & 50 & 28 & 28.992 & 17--26 & 3 & 1086 \\ \hline
4 & 10096 & 0.022 & 6 & 32 & 7 & 7.222 & 4--7 & 3, 5, 2, 1, 7 & 10 \\ \hline
5 & 20889 & 0.046 & 4 & 49 & 7 & 8.76 & 1--10 & 2 & 82 \\ \hline
6 & 5268 & 0.012 & 5 & 25 & 5 & 6.081 & 2, 4, 5 & 1--7 & 7 \\ \hline
7 & 54575 & 0.120 & 9 & 42 & 12 & 12.286 & 6--13 & 3, 2 & 131 \\ \hline
8 & 8511 & 0.019 & 4 & 27 & 4 & 4.47 & 1--3 & 1--5, 7 & 1 \\ \hline
9 & 277118 & 0.609 & 4 & 42 & 5 & 5.764 & 1--4 & 1--7 & 349 \\ \hline
10 & 18943 & 0.042 & 4 & 39 & 6 & 7.052 & 1--8 & 5, 1 & 78 \\ \hline
\end{tabular}
\end{center}
}
\end{table}

In Table \ref{Tab:stats_for_NMF_clustering}, column `Active quarter statistics' represents the minimum, maximum, median or average number of quarters that the group of editors stay active in Wikipedia; column `Dominant user roles' indicates which user roles are dominant in the profiles for each cluster as suggested by NMF clustering. Note that user role 3 is one of the dominant features in 8 clusters, suggesting that user role 3 is a  common and important role in Wikipedia. Column `\#admins' represents the number of administrators in the clusters: Cluster 3 contains the largest number of administrators. We observe from Table \ref{Tab:stats_for_NMF_clustering} that Cluster 9 is the largest cluster with about 61\% of editors: its average number of active quarters is about 6, the editors in this cluster mainly active in quarter 1--4, suggesting that a large proportion of editors in this cluster stayed active in Wikipedia for only 4 quarters. Cluster 6 is the smallest cluster with only 1.2\% of editors: its dominant quarters is 1--6, indicating that this group of editors also stayed active for relatively short periods of time compared with other clusters. The editors in Cluster 2 and 3 stayed active for very long periods of time, with their average number of active quarters being 18 and 28, respectively. As we will show next, the NMF clustering provides a reasonable good clustering of editor profiles.

\begin{figure}[!htbp]
{\fontsize{10.0pt}{10.0pt}\selectfont
\begin{center}
  \subfloat[Cluster 3]{\includegraphics[width=.95\textwidth]{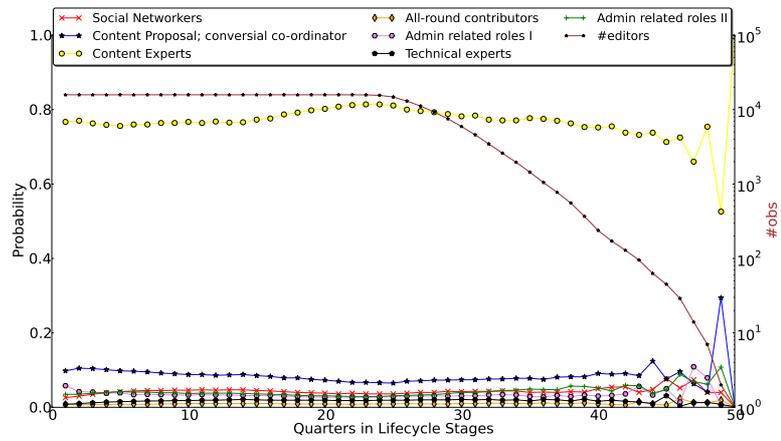}}\\
\vspace{-.8em}
  \subfloat[Cluster 5]{\includegraphics[width=.95\textwidth]{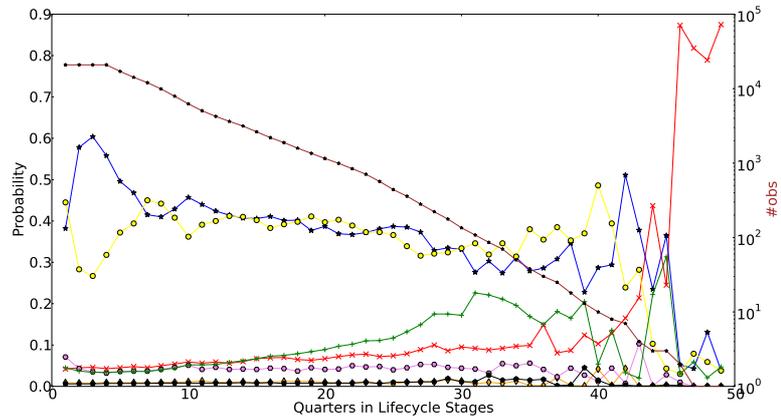}}\\

  \subfloat[Cluster 8]{\includegraphics[width=.85\textwidth]{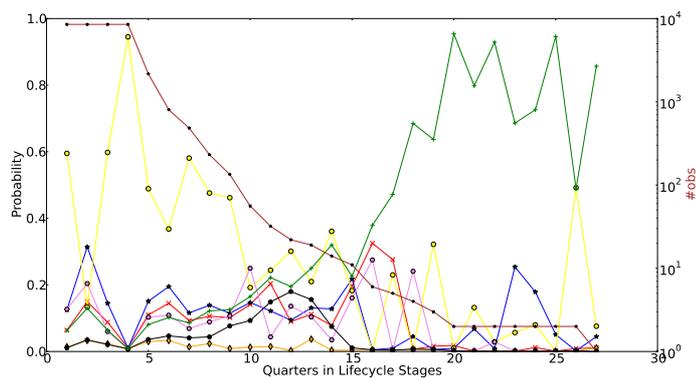}}
\end{center}
}
\vspace{-.8em}
\caption{Lifecycle on a group basis: x-axis represents the relative quarter in editors' lifestage, the left y-axis represents the  the probability of a user role being assigned to editor profile, the right y-axis represents the number of editor profiles observes in a quarter. Different curves represent the evolution of profile assignments.}
\vspace{-.8em}
\label{fig:group_profile_evolution}
\end{figure}

\textbf{Group-level change in editor profiles.} We now rely on the clustering of editors to analyze the general development of editor profiles on a group basis. Following previous studies \cite{Danescu-Niculescu-Mizil2013,Rowe2013} in which the general evolution of user lifecycle was visualized against user life-stage, Figure \ref{fig:group_profile_evolution} plots the average assignment of each editor profile to user roles at different life-stages for 3 clusters\footnote{The plots for Cluster 1, 4, 6 are very similar to that of Cluster 8; the trend of the plots for Cluster 2, 7, 9 is similar to that of Cluster 3; the plot for Cluster 10 is very similar to that of Cluster 5; these plots are omitted due to space limitation.}. Inspecting the evolution of profile assignments over the editors' life-stages in Figure \ref{fig:group_profile_evolution}, we make the following observations for two group of editors:
\vspace{-.5em}
\begin{itemize}
	\item Based on the trend of evolution in the plots and the statistics in Table \ref{Tab:stats_for_NMF_clustering}, we can further divide the editor profiles into 2 groups: long term editors with their profile assignments experienced relatively soft evolution (Cluster 2, 3, 5, 7, 9, 10), short term editors with their profile assignments experienced considerably fluctuated evolution (Cluster 1, 4, 6, 8). On average, group of long term editors stay active in Wikipedia for much longer time periods than group of short term editors. 
	\item Long term editors: in the early stage of lifespans, they generally focused their contribution in one or two namespaces (e.g., \emph{main} namespace in Cluster 2, 3, 7, 9; \emph{main} and \emph{article talk} namespaces in Cluster 5; \emph{main} and \emph{user} namespaces in Cluster 10); in the late stage of lifespans, this group of editors gradually diversified their participation by shifting edit preference from dominant namespaces to multiple namespaces; we also observe from Figure \ref{fig:group_profile_evolution}(a) that editors in Cluster 3 mainly edited the \emph{main} namespace over the course of their career.
	\item Short term editors: their profile assignments experienced more fluctuations across their lifecycle periods, indicating these users did not develop long term edit interest in one namespace and generally distributed their contribution randomly among the namespaces.
\end{itemize}

To summarize, we find that: long term editors generally have long term edit preference in one or more namespances throughout their career and diversify their participation in the late stage of their lifespans; by contrast, short term editors generally do not develop long term edit interest and tend to distribute edits among the namespaces, thus experience more fluctuation in their profile assignments. The observations suggest that the fluctuation in profile assignments may be an indicator that users are likely to leave the community. Next, we generate features corresponding to these findings for churn prediction task.

\vspace{-3 mm}
\section{Application of Editor Lifecycles to Churn Prediction}
\label{sec:Churn_prediction}
We have so far focused on understanding the lifecycle trajectories of editors based on their exhibited behavior. We now move on to exploring how predictive are the features generated from patterns of change in editor profile in identifying whether a user will abandon a community. Churners present a great challenge for community management and maintenance as the leaving of established members can have a detrimental effect on the community in terms of creating communication gap, knowledge gap or other gaps. 

\textbf{Definition of churn prediction.} Following the work by Danescu-Niculescu-Mizil {\it et al.} \cite{Danescu-Niculescu-Mizil2013}, we define churn prediction task as predicting whether an editor is among the `departed' or the `staying' class. Considering that our dataset spans for about 13 years (i.e. 50 quarters), and that studies about churn prediction generally follow the diagram of predicting the churn status of users in the prediction period based on user exhibited behavior in the observation period (e.g. \cite{Weia2002,Danescu-Niculescu-Mizil2013}), we employed sliding-window based method for churn prediction. Specifically, we make predictions based on features generated from editor profile assignments in a $w$=4 quarters sliding window. An editor is in the `departed' class if she leaved the community before active for less than $m$=1 quarter after the sliding window, lets denote the interval [w, w+m] as the departed range. Similarly, an editor is in the `staying' class if she stayed active in the community long enough for a relatively large $n\geq 3$ quarters after the sliding-window; we term the interval [$w+n$, $+\infty$] as the staying range.

\textbf{Features used for the task.} Our features are generated based on the findings reported in the previous section. For simplicity, we assume the $w$ quarters included in the $i$-th sliding-window being $i$=$[j,\cdots,j+w-1$] ($j\in[1,50]$), and denote the Probability Of Activity Profile of an editor in quarter $j$ being assigned to the $k$-th user role as $\textrm{POAP}_{i,j,k}$. We use the following features to characterize the patterns of change in editor profile assignments:
\begin{itemize}
	\item {\it First active quarter}: the quarter in which an editor began active in Wikipedia. The timestamp a user joined the community may affect her decision about whether to stay for longer as research suggested that users joined later in Wikipedia may face a more severe situation in terms of contribution being accepted by the community and being excluded by established members.
	\item {\it Cumulative active quarters}: the total number of quarters an editor had been active in the community till the last quarter in the sliding window.
	\item {\it Fraction of active quarters in lifespan}: the proportion of quarters a user stayed active till the sliding window. For instance, if an editor joined in quarter 10, and stayed active for 8 quarters till the current sliding window (16--19), then the figure is calculated as: 8/(19-10+1) = 0.80. 
	\item {\it Fraction of active quarters in sliding window}: the fraction of quarters a user stayed active in current sliding window. 
	\item {\it Diversity of edit activity}: denotes the average entropy of $\textrm{POAP}_{i,j,k}$ for each quarter $j$ in window $i$. This measure captures the extent to which an editor diversified her edit towards multiple namespaces.
	\item {\it mean($\textrm{POAP}_{i,j,k}$)}: denotes the average of $\textrm{POAP}_{i,j,k}$ for each user role $k$ in window $i$. This measure captures whether an editor focused on one or more namespace in window $i$.
	\item {\it $\Delta \textrm{POAP}_{i,j,k}$}: denotes the change in $\textrm{POAP}_{i,j,k}$ between the quarter $j-1$ and $j$ (for $j\in[2,50]$), and is measured by $\Delta \textrm{POAP}_{i,j,k}$=$ (\textrm{POAP}_{i,j,k} - \textrm{POAP}_{i,j-1,k} + \delta)/(\textrm{POAP}_{i,j-1,k} + \delta)$, where $\delta$ is a small positive real number (i.e. 0.001) to avoid the case when $\textrm{POAP}_{i,j-1,k}$ is 0. This measure also captures the fluctuation of $\textrm{POAP}_{i,j,k}$ for each user role $k$ in window $i$. 
\end{itemize}

For each editor, the first three features are global-level features which may be updated with the sliding window, the remaining four features are window-level features and are recalculated within each sliding window. The intuition behind the last four features is to approximate the evolution of editor lifecycle we sought to characterize in previous section. The dataset is of the following form: $D$=${(x_{i},y_{i})}$, where $y_{i}$ denotes the churn status of the editor, $y_{i}\in\{\textrm{Churner, Non-churner}\}$; $x_{i}$ denotes the feature vector for the editor.  

\textbf{Experimental setup.} The prediction task is a binary classification problem, we use the RandomForest algorithm\footnote{We also experimented with other algorithms (e.g. Logistic regression and SVM) and obtained similar performance.} for the purpose. To avoid bias in the results due to highly imbalance of class distribution, for each sliding window, we randomly sampled the editors in order to generate a dataset with a desired class ratio (churners:non-churners) being 1:2. Each time we slide the window by one quarter. The results reported next are averaged over 10-fold cross-validation.

\begin{figure}[!htb]
\baselineskip=12pt
{\fontsize{10.0pt}{10.0pt}\selectfont
\begin{center}
  \subfloat{\includegraphics[width=1.0\textwidth]{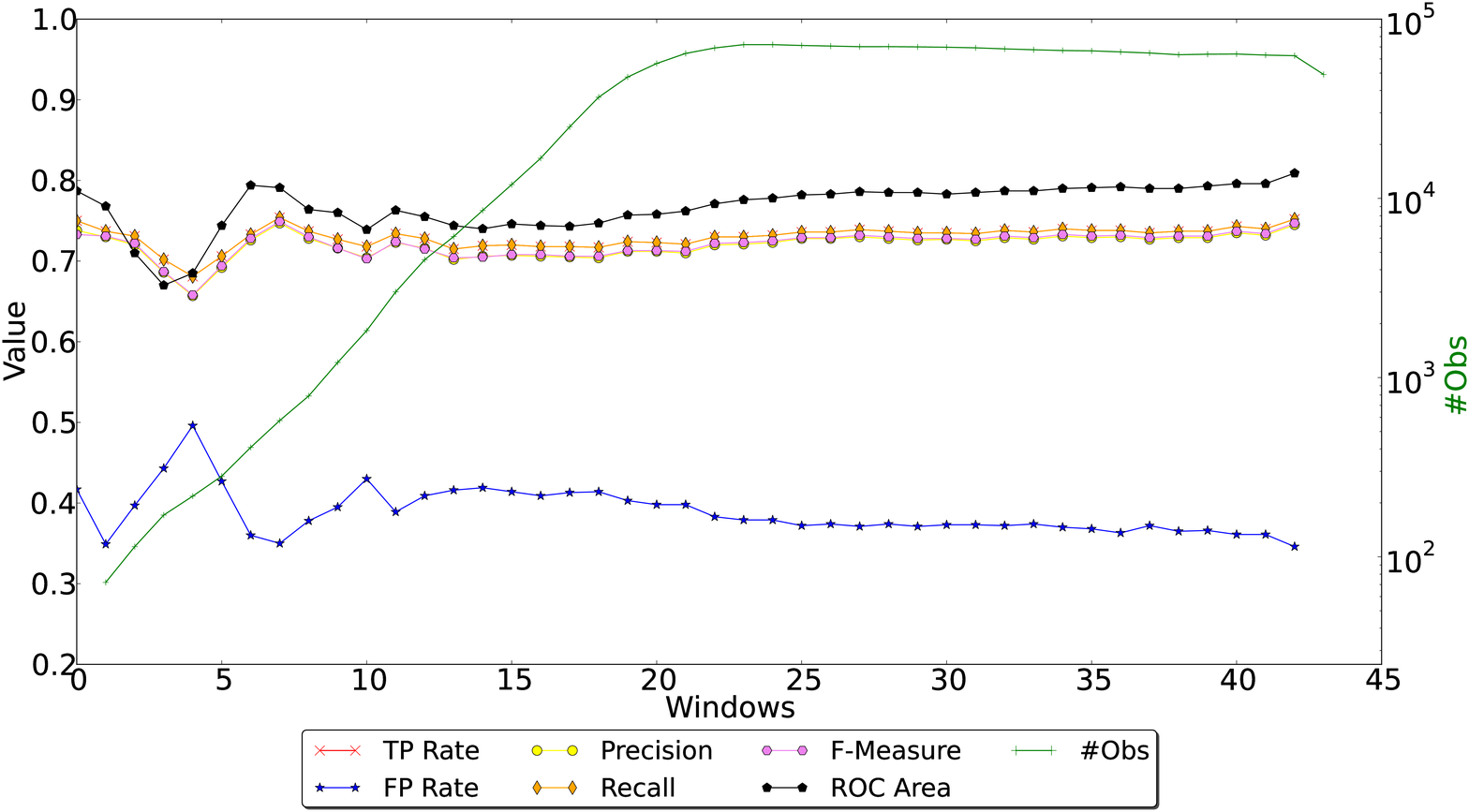}}\quad
\vspace{-.8em}
\caption{Performance for sliding-window based churn prediction. X-axis represents the sliding window, the left y-axis represents the performance, the right y-axis represents the number of editor profiles observes in a window.}
\vspace{-1.2em}
\label{fig:Churn_prediction_performance}
\end{center}
}
\end{figure}

\textbf{Performance of sliding-window based churn prediction.} Figure \ref{fig:Churn_prediction_performance} plots the performance of churn prediction. We observe that in the first 10 windows, the number of editors observes in each window is relatively small (less than 1000), which results in a fluctuation in all performance measures; from the 10th window onwards, the number of editors grows steadily and then maintains at a level of about 65,000 after the 20th window, the performance measures become relatively stable: with \textsf{FP rate} being \textbf{0.31}, \textsf{ROC area} being \textbf{0.80}, other measures (i.e. \textsf{TP rate}, \textsf{Precision}, \textsf{Recall}, \textsf{F-measure}) being \textbf{0.75}.

\vspace{-8 mm}
\vspace{-3 mm}

Note when Danescu-Niculescu-Mizil {\it et al.} \cite{Danescu-Niculescu-Mizil2013} performed churn prediction on two beer rating communities (i.e. BeerAdvocate and RateBeer) based on user's linguistic change features, they obtained the best performance of \textsf{Precision} being 0.77, \textsf{Recall} being 46.9, \textsf{F-measure} being 0.56. Rowe \cite{Rowe2013} evaluated churn prediction task on three online platforms (i.e. Facebook, SAP and Server Fault) using user's social and lexical change related features, and obtained the best performance of \textsf{Precision@K} being 0.791 and \textsf{AUC} being 0.617. Comparison of the figures indicates that our study achieves at least comparable overall and average performance with the other two studies for churn prediction in online communities. This observation suggests that in online communities, the sudden change in user activity can be an important signal that the user is likely to abandon the community.

\begin{figure}[!htb]
\baselineskip=12pt
{\fontsize{10.0pt}{10.0pt}\selectfont
\begin{center}
  \subfloat{\includegraphics[width=1.0\textwidth]{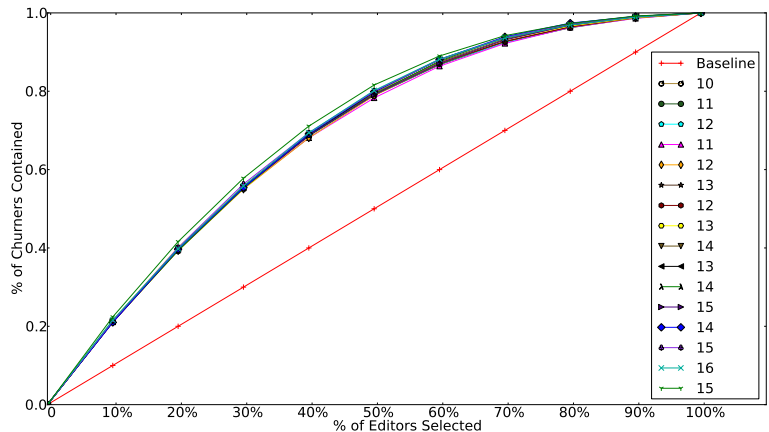}}\quad
\vspace{-.8em}
\caption{Lift chart obtained by the proposed churn-prediction model. Different curves represent the lift curves for different sliding windows.}
\vspace{-1.2em}
\label{fig:Churn_prediction_cumulative_gains}
\end{center}
}
\end{figure}

\textbf{Cumulative gains for churn prediction.} The lift factors are widely used by researchers to evaluate the performance of churn-prediction models (e.g. \cite{Weia2002}). The lift factors achieved by our model are shown in Figure \ref{fig:Churn_prediction_cumulative_gains}. In lift chart, the diagonal line represents a baseline which randomly selects a subset of editors as potential churners, i.e., it selects s\% of the editors that will contain s\% of the true churners, resulting in a lift factor of 1. In Figure \ref{fig:Churn_prediction_cumulative_gains}, on average, our model was capable of identifying 10\% of editors that contained 21.2\% of true churners (i.e. a lift factor of 2.12), 20\% of editors that contained 39.3\% of true churners (i.e. a lift factor of 1.97), and 30\% of editors that contained 54.7\% of true churners (i.e. a lift factor of 1.82). Evidently, our model achieved higher lift factors than the baseline. Thus if the objective of the lift analysis is to identify a small subset of likely churners for an intervention that might persuade them not to churn, then this analysis suggests that our model can identify a set of 10\% of users where the probability of churning is more than twice the baseline figure.

\vspace{-8 mm}
\vspace{-3 mm}

\textbf{Feature analysis.} To better understand the contribution of each feature set in the context of all the other features, we performed an ablation study. Specifically, for each set of features, we ran a classifier to contain all of the features except the target set. We then compared the performance of the resulting classifier to that of classifier with complete set of features. We find that the classifier with $\Delta \textrm{POAP}_{i,j,k}$ excluded results in the most decrease in performance, indicating that the sudden change in user behavior can be an important signal that the user is likely to abandon the community. We also observe that none of the features can by themselves achieve the performance reported in Figure \ref{fig:Churn_prediction_performance}, which suggests that the features we generated complement each other in predicting churners in online communities.

\vspace{-0.08in}
\section{Conclusion}
\vspace{-0.08in}
In this paper we have presented a novel latent space analysis of editor lifecycles in online communities based on user exhibited behavior. The analysis reveals a number of different categories of editor (e.g. content experts, social networkers) and provides a means to track the evolution of editor behavior over time. We also perform a non-negative matrix factorization clustering of editor profiles in the latent space representation and find that long term and short term users generally have very different profiles and evolve differently in their lifespans. 

We show that understanding patterns of change in user behavior can be of practical importance for community management and maintenance, in that the features inspired by our latent space analysis can differentiate churners from non-churners with reasonable performance. This work opens a few interesting questions for future research. Firstly, social networking plays a foundational role in online communities for knowledge and resource sharing and member retention, future work should accommodate social dynamics in the model. Secondly, the topic model we used in this paper did not account for the evolution of activity in user-level, it will be very helpful to develop topic models that accommodate the evolution in the user and community level. Moreover, online platforms are very similar in terms of allowing multiple dimensions of user activities, the approach presented in this paper can be generalized to other platforms very easily.

\vspace{3 mm}\noindent{\emph{Acknowledgements.}}
This work is supported by Science Foundation Ireland (SFI) under Grant Number SFI/12/RC/2289 (Insight Centre). Xiangju Qin is funded by UCD-CSC Scholarship Scheme 2011.

\vspace{-0.20in}
\small{
\bibliographystyle{splncs03}
\bibliography{wiki_user_lifecycles_ref}

\begin{thebibliography}{10}
\providecommand{\url}[1]{\texttt{#1}}
\providecommand{\urlprefix}{URL }

\bibitem{Ahmed2010}
{Ahmed}, A., {Xing}, E.P.: {Timeline: A Dynamic Hierarchical Dirichlet Process
  Model for Recovering Birth/Death and Evolution of Topics in Text Stream}. In:
  Proceedings of UAI 2010. pp. 20--29. AAAI (2010)

\bibitem{Blei2006}
Blei, D.M., Lafferty, J.D.: {Dynamic Topic Models}. In: Proceedings of ICML'06.
  pp. 113--120. ACM (2006)

\bibitem{Blei2003}
{Blei}, D.M., {Ng}, A.Y., {Jordan}, M.I.: {Latent Dirichlet Allocation}.
  Journal of Machine Learning Research  3,  993--1022 (2003)

\bibitem{bout08headstart}
Boutsidis, C., Gallopoulos, E.: {SVD based initialization: A head start for
  non-negative matrix factorization}. Pattern Recognition  (2008)

\bibitem{Chan2010}
{Chan}, J., {Hayes}, C., {Daly}, E.M.: {Decomposing Discussion Forums using
  User Roles}. In: Proceedings of ICWSM 2010. pp. 215--218 (2010)

\bibitem{Danescu-Niculescu-Mizil2013}
{Danescu-Niculescu-Mizil}, C., {West}, R., {Jurafsky}, D., {Leskovec}, J.,
  {Potts}, C.: {No Country for Old Members: User Lifecycle and Linguistic
  Change in Online Communities}. In: Proceedings of WWW'13. pp. 307--318. Rio
  de Janeiro, Brazil (2013)

\bibitem{Furtado2013}
Furtado, A., Andrade, N., Oliveira, N., Brasileiro, F.: {Contributor Profiles,
  their Dynamics, and their Importance in Five Q\&A Sites}. In: Proceedings of
  CSCW'13. pp. 1237--1252. ACM, San Antonio, Texas, USA (2013)

\bibitem{lin07gradient}
Lin, C.: {Projected gradient methods for non-negative matrix factorization}.
  Neural Computation  19(10),  2756--2779 (2007)

\bibitem{Panciera2009}
Panciera, K., Halfaker, A., Terveen, L.: Wikipedians are born, not made: A
  study of power editors on wikipedia. In: Proceedings of GROUP'09. pp. 51--60.
  ACM (2009)

\bibitem{Rowe2013}
{Rowe}, M.: {Mining User Lifecycles from Online Community Platforms and their
  Application to Churn Prediction}. In: Proceedings of ICDM 2013. pp. 1--10.
  IEEE, Dallas, Texas, USA (2013)

\bibitem{Wang2006}
Wang, X., McCallum, A.: {Topics over Time: A non-Markov Continuous-time Model
  of Topical Trends}. In: Proceedings of KDD'06. pp. 424--433. ACM (2006)

\bibitem{Weia2002}
{Weia}, C.P., {Chiub}, I.T.: {Turning telecommunications call details to churn
  prediction: a data mining approach}. Expert Systems with Applications  23(2),
   103--112 (2002)

\bibitem{Welser2011}
Welser, H.T., Cosley, D., Kossinets, G., Lin, A., Dokshin, F., Gay, G., Smith,
  M.: Finding social roles in wikipedia. In: Proceedings of iConference'11. pp.
  122--129. ACM (2011)

\end{thebibliography}
}

\end{document}